\documentclass{llncs}

\usepackage[table,xcdraw]{xcolor}
\usepackage[english]{babel}
\usepackage[utf8x]{inputenc}
\usepackage{amsmath}
\usepackage{todonotes}
\usepackage[hyphens]{url}
\usepackage{hyperref}
\usepackage{enumitem}
\usepackage{booktabs}
\usepackage{nicefrac}
\usepackage{multirow}
\usepackage{makecell}
\usepackage{cite}
\usepackage[flushleft]{threeparttable}
\usepackage{pgfplots}
\usepackage[export]{adjustbox} % subfloat vertical alignment (https://tex.stackexchange.com/questions/296624/subfloat-vertical-alignment-in-latex)
\usepackage{subfig}
\pgfplotsset{compat=1.5} % tikzpicture: avoid overlap of yLabel with y acis (https://tex.stackexchange.com/questions/46634/ylabel-postion-with-pgfplots)

\newcommand\eventTL[0]{EventKG+TL}

\begin{document}

%\title{\eventTL{}: Adopting a Multilingual Event-Centric Knowledge Graph for Timeline Generation}
\title{\eventTL{}: Creating Cross-Lingual Timelines from an Event-Centric Knowledge Graph}

\author{Simon Gottschalk \and Elena Demidova}

\institute{L3S Research Center, Leibniz Universit\"at Hannover, Hannover, Germany \\
 \email{\{gottschalk, demidova\}@L3S.de}
 } 

\maketitle    

\begin{abstract}
The provision of multilingual event-centric temporal knowledge graphs such as EventKG enables structured access to representations of a large number of historical and contemporary events in a variety of language contexts. 
Timelines provide an intuitive way to facilitate an overview of events related to a \textit{query entity} - i.e. an entity or an event of user interest - over a certain period of time.
In this paper, we present \eventTL{} - a novel system that generates cross-lingual event timelines using EventKG and facilitates an overview of the language-specific event relevance and popularity along with the cross-lingual differences. 

\end{abstract}

\textbf{Demo URL: \url{http://eventkg.l3s.uni-hannover.de/eventkg_tl}}

\section{Introduction}
\label{sec:introduction}

The amount of event-centric information regarding contemporary and historical events of global importance, such as Brexit and the migration crisis in Europe, constantly grows on the Web, in Web archives, in the news as well as within emerging event-centric collections \cite{GossenDR15} and knowledge graphs generated from these sources (e.g. \cite{gottschalk2018eventkg}, \cite{ROSPOCHER2016132}). 
An important research area in this context is cross-cultural and cross-lingual event analytics (e.g. see \cite{Rogers:2013}, \cite{GottschalkDBR17} for case studies, and \cite{GottschalkD17} for a cross-lingual user interface). These studies aim to analyze language-specific and community-specific representations and perceptions of historical and contemporary events including their popularity and relations in a language context as well as to better understand the cross-lingual differences.

EventKG \cite{gottschalk2018eventkg} - a recently proposed multilingual event-centric temporal knowledge graph incorporating over $690$ thousand events in five languages - is an important knowledge source that can facilitate a variety of studies and applications related to cross-cultural and cross-lingual event analytics.  
However, given a \textit{query entity}, i.e. an entity or 
an event of user interest, EventKG can contain hundreds of related events along with their descriptions in several language contexts, which makes the provision of a comprehensive cross-lingual overview and a selection of relevant events for further detailed analysis challenging.

Timelines are an intuitive way to provide an overview of events related to a \textit{query entity} over a certain period of time. 
Timeline generation is an active research area \cite{Althoff:2015}, where the focus is to generate a timeline (i.e. a chronologically ordered selection) of events related to the \textit{query entity} from a knowledge graph.
However, existing timelines do not explicitly support a cross-lingual comparison of language-specific event representations, including their popularity and relation to the \textit{query entity} in different language contexts.  

\eventTL{} presented in this paper is a timeline generator that creates cross-lingual timelines for a \textit{query entity}, while relying on EventKG to provide language-specific information with respect to the event popularity and the relation strength between the events and the \textit{query entity}.
To this extent, \eventTL{} conducts a language-specific event ranking and complements this ranking with a cross-lingual visual representation.
The timelines generated by \eventTL{} facilitate efficient identification of relevant events based on their language-specific popularity, relation strength and the cross-lingual differences. 

\section{Scenarios \& Timelines}
\label{sec:scenarios}

A \emph{multilingual event-centric temporal knowledge graph} 
$kg=(L, E, R)$ is a labeled directed multigraph, where
$L$ is a set of language contexts, 
$E$ is a set of nodes (i.e. events or entities), and   
$R$ is a multiset of directed edges (i.e. relations).  

Given a \textit{query entity} $q \in E$, the timelines generated by \eventTL{} can assist users in answering questions such as: 

\textit{$Q_1$: What are the most popular events related to $q$?}

\textit{$Q_2$: Which events are the most closely related to $q$?}

\textit{$Q_3$: Which of the most popular events are the most closely related to $q$?}

\textit{$Q_4$: How does the popularity 
of the identified events and the strength of their relations to the \textit{query entity} $q$ differ across the language contexts?}

The provision of \eventTL{} facilitates users to answer these questions with respect to a particular language context $l \in L$ and enables a visual cross-lingual comparison.
To answer these questions, the user of \eventTL{} can issue 
a \textit{timeline query} that includes the following parameters:

\begin{itemize}
\item a \textit{query entity} $q \in E$;
\item a set of the language contexts of user interest $L' \subseteq L$;
\item the maximum number $k$ of the events to be selected per language context; 
\item the ranking criterion $rc_{i}$ to identify the top-$k$ most relevant events among all events $E' \subset E$ related to $q$ in $kg$ according to the questions $Q_1-Q_3$. 
\end{itemize}

The ranking criteria include: 
\begin{itemize}
\item [$rc_{1}$:] $popularity(e, l)$ is the popularity of an event $e\in E'$ in $l\in L'$; 
\item [$rc_{2}$:] $relation~strength(q,e,l)$ is the relation strength between the \textit{query entity} $q$ and an event $e \in E'$ in a language context $l\in L'$; and 
\item [$rc_{3}$:] $combined(q,e,l)$ is a combination of the event popularity of $e \in E'$ and the relation strength between $e$ and the \textit{query entity} $q$ in $l\in L'$.
\end{itemize}

The timelines generated by \eventTL{} complement the language-specific event ranking with a cross-lingual visual representation 
to address the question $Q_4$. To this extent, \eventTL{} utilizes labeled pie charts located on a timeline, where each pie chart represents an individual event.  
The size of the pie chart corresponds to an overall (i.e. language independent) relevance of the event according to the ranking criterion $rc_{i}$.
Each slice of the pie chart represents a language context. The area of each slice is proportional to the contribution of the corresponding language context to the ranking criterion $rc_{i}$. 

\begin{figure}[!t]
 \centering
  \includegraphics[width=\textwidth]{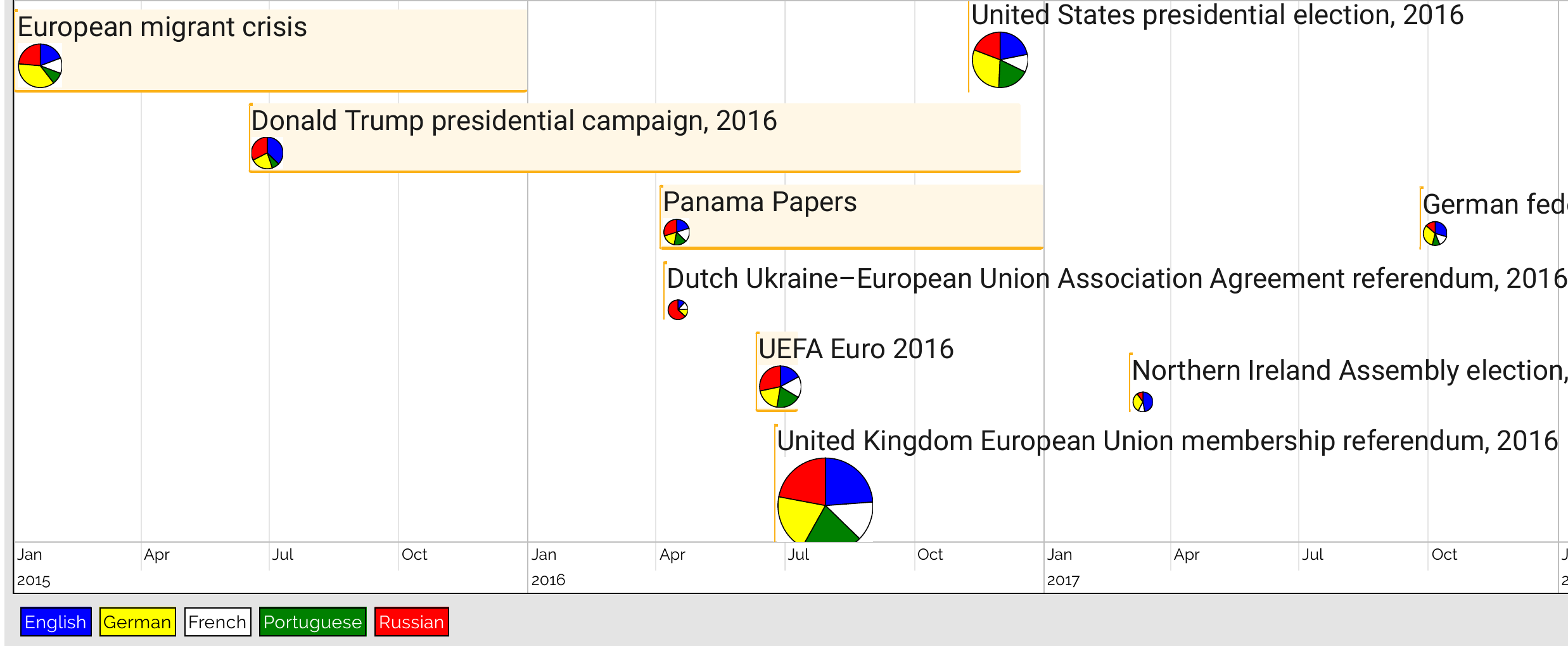}
  \caption{An excerpt of an \eventTL{} timeline representing events related to the \textit{query entity} "Brexit" in the time interval 01/2015-01/2018, overall including the top-8 events from each of the five language contexts in EventKG ranked according to $rc_{3}$ - i.e. a combination of the popularity and the relation strength of the events towards Brexit. 
  Each event is represented by a labeled pie chart. The size of the pie chart corresponds to the language independent event relevance according to \textbf{$rc_{3}$}. 
  The colored slices determine the ratio of the relevance in a language context (see the legend for the color encoding). The duration of events that lasted for more than a day is marked by a yellow interval. Upon click on a timeline entry, detailed information, including scores and link counts, is shown.
  }
  \label{fig:timeline}
\end{figure}

Fig. \ref{fig:timeline} exemplifies a Brexit timeline. We can observe that the
most important event according to $rc_{3}$ is the 
\textit{"United Kingdom European Union membership referendum, 2016"} that is nearly equally important in all considered language contexts. 
Some of the events are more important in the specific language contexts, e.g.
\textit{"European Migrant Crisis"} in the German and 
\textit{"Dutch Ukraine-European Union Association Agreement referendum 2016"} in the Russian context.

\section{Timeline Generation}
\label{sec:creation}

\subsubsection{The Knowledge Graph}
To answer a \textit{timeline query}, \eventTL{} utilizes EventKG \cite{gottschalk2018eventkg}. 
EventKG is a multilingual RDF knowledge graph incorporating over $690$ thousand events and over $2.3$ million temporal relations in V1.1 extracted from several large-scale entity-centric knowledge graphs (i.e. Wikidata, DBpedia in five language editions and YAGO), Wikipedia Current Event Portal (WCEP) and Wikipedia event lists. 
One of the key features of EventKG is the provision of event-centric information for historical and contemporary events, including their interlinking in the language-specific contexts to facilitate an assessment of relation strength 
and event popularity. 
The information on language-specific interlinking  provided by EventKG is based on the corresponding Wikipedia language editions.

\subsubsection{Event and Relation Retrieval}
To retrieve relevant information from EventKG, \eventTL{} adopts SPARQL queries.
First, \eventTL{} retrieves the \textit{query entity} $q$, including its existence time, if available. Second, \eventTL{} retrieves a set of events $E'\subset E$ that are connected to $q$ via an EventKG relation as the subject or the object, along with the time information associated with these events. Third, the interlinking information related to the events in $E'$ is retrieved from EventKG's link relations and their \texttt{eventKG-s:links} and \texttt{eventKG-s:mentions} property values.

\subsubsection{Event Ranking and Timeline Creation}
The top-$k$ events related to $q$ are selected according to the ranking criterion. For each event $e \in E'$ and language $l \in L'$, the language-specific relevance score is computed using the interlinking information provided by EventKG. The following link counts are used:

\begin{itemize}
\item $count_{links}(e,l)$: Event link count, i.e. the number of links pointing to the event $e$ in a language context $l$ (via \texttt{eventKG-s:links}).
\item $count_{pair}(q,e,l)$: Pair count, i.e. the number of links from $q$ to $e$ plus the number of links from $e$ to $q$ in $l$, denoted by \texttt{eventKG-s:links} values.
\item $count_{mentions}(q,e,l)$: Mention count, i.e. the number of sentences in a language context $l$ that jointly link to $q$ and $e$, denoted by \texttt{eventKG-s:mentions}.
\end{itemize}

Each count is normalized to $[0,1]$ by dividing its value by the highest value of this count related to the events in $E'$ in the respective language. That way, the bias resulting from the differences in the language-specific coverage is reduced. 
To avoid the domination of the disproportionately often linked events (e.g. the World War II), a smoothing parameter $\alpha$, experimentally set to $0.25$, is adopted. 
The scores are computed as follows:

\begin{equation}
\text{popularity}(e,l) = \left(\frac{count_{links}(e,l)}{max \{ count_{links}(e',l) | e' \in E' \}}\right) ^ \alpha
\end{equation}

\begin{align}
\begin{split}
\text{relation strength}(q,e,l) = &\ \frac{1}{2} \cdot \left(\frac{count_{pair}(q,e,l)}{max \{ count_{pair}(q,e',l) | e' \in E' \}}\right) ^ \alpha \\
& + \frac{1}{2} \cdot \left(\frac{count_{mentions}(q,e,l)}{max \{ count_{mentions}(q,e',l) | e' \in E' \}}\right) ^ \alpha
\end{split}
\end{align}

The $combined$ score ($rc_3$) is computed as a linear combination of the two ranking criteria. We experimentally set its weight to $w = \nicefrac{1}{3}$.

\begin{align}
\begin{split}
\text{combined}(q,e,l) = &\ w \cdot \text{popularity}(e,l) \\
& + (1 - w) \cdot \text{relation strength}(q,e,l)
\end{split}
\end{align}

The resulting timeline consists of a chronologically ordered list of the top-$k$ highest ranked events per language with respect to the ranking criterion.

\subsubsection{System Implementation}

The \eventTL{} system is accessible as an HTML5 website. 
It is implemented using the Java Spark web framework\footnote{\url{http://sparkjava.com/}}. The timeline is visualized through the browser-based Javascript library vis.js\footnote{\url{http://visjs.org/timeline_examples.html}}, the pie charts are created using the Google Charts Javascript library\footnote{\url{https://developers.google.com/chart/interactive/docs/gallery/piechart}} and pop-ups showing detailed event information are based on Twitter Bootstrap\footnote{\url{https://getbootstrap.com/}}. 

\section{Demonstration}
\label{sec:demonstration}

In our demonstration we will primarily show how
\eventTL{} works and how users can use it to 
create cross-lingual timelines. 
To highlight the advantages of our approach, we will ask our audience to create timelines for the entities and events of their choice using \eventTL{} based on the language-specific information contained in EventKG. 
Through the visual cross-lingual comparison provided by \eventTL{}, the audience can get an impression 
of the language-specific event representations, as well as their relation to the \textit{query entity} and popularity in different language contexts.

\bibliographystyle{splncs03}

{\footnotesize

\subsubsection*{Acknowledgements} This work was partially funded by  the ERC ("ALEXANDRIA", 339233) and BMBF ("Data4UrbanMobility", 02K15A040).

\bibliography{references}}

\end{document}